\documentclass[aps,prd,reprint,showkeys,nofootinbib]{revtex4-1}
\usepackage[utf8]{inputenc}
\usepackage{amsmath}
\usepackage{amsfonts}
\usepackage{amssymb}
\usepackage{hyperref}
\usepackage{color}
\usepackage{graphicx}
\usepackage{epstopdf}

\begin{document}

\title{Out-of-equilibrium chiral magnetic effect and momentum relaxation in holography}

\author{Jorge Fern\'andez-Pend\'as}
\email{j.fernandez.pendas@csic.es}
\author{Karl Landsteiner}
\email{karl.landsteiner@csic.es}

\affiliation{Instituto de F\'isica Te\'orica UAM/CSIC, c/Nicol\'as Cabrera 13-15, Universidad Aut\'onoma de Madrid, Cantoblanco, 28049 Madrid, Spain}
\preprint{IFT-UAM/CSIC-19-105}
\date{\today}

\begin{abstract}
We compute anomalous transport phenomena sourced by vector and axial magnetic fields in out-of-equilibrium setups produced by Vaidya background metrics in holography. We use generalized Vaidya metrics that include momentum relaxation induced by
massless scalar fields. While the background metric and gauge field show formally instantaneous thermalization the chiral magnetic effect has significantly large equilibration times. We study how the equilibration of the chiral magnetic effect depends on the length of the Vaidya quench and the momentum relaxation parameter. 
These results shed some light on aspects of the chiral magnetic effect in out-of-equilibrium situations such as the quark gluon plasma produced in heavy ion collisions.
\end{abstract}

\keywords{Anomalies, Transport Phenomena, Out-of-Equilibrium}

\maketitle

\section{Introduction}\label{sec:intro}

Anomaly-induced transport phenomena have been the subject of much interest (see Refs. \cite{Kharzeev:2013ffa,Landsteiner:2016led} for reviews). The most common phenomenon is the so-called chiral magnetic effect (CME) \cite{Fukushima:2008xe}. It gives rise to the spontaneous generation of a current in the presence of a chiral imbalance and a background magnetic field. It is a direct consequence of the axial anomaly. An important aspect of the CME is that in a subtle way it is always related to some non-equilibrium physics. On the theoretical level the CME in a gauge current, such as the electric current, has to vanish in strict equilibrium. This is known as Bloch theorem and a discussion in relation to the CME has been given in Ref. \cite{Yamamoto:2015fxa}. A careful examination of the CME shows that it indeed vanishes in equilibrium due to a topological contribution from a counterterm that arises in the definition of the electric current \cite{Gynther:2010ed}. Currents related to anomalous global symmetries, in contrast, can have non vanishing expectation values. 

The nonequilibrium aspect is not only important in theory. It is also essential in experimental situations in which the CME arises, like condensed matter systems known at Weyl semimetals \cite{Li:2014bha} or the quark gluon plasma in heavy ion collisions. In Weyl semimetals non-equilibrium arises due to the application of an electric field parallel to the magnetic field. The CME induces then the celebrated negative magnetoresistivity \cite{Son:2012bg}. In heavy ion collisions the chiral imbalance is induced in the early far-from-equilibrium stages by the gluonic contribution to the axial anomaly \cite{Kharzeev:2007jp}.

So far, most of the theoretical investigations of anomaly-induced transport have concentrated on equilibrium or near-equilibrium situations which can be described by hydrodynamics \cite{Son:2009tf}. The considerations above make it clear, however, that a much better understanding of anomaly induced transport out-of-equilibrium is needed. 

Holography not only is by now a standard tool  to investigate transport in strongly coupled systems, such as the quark gluon plasma, but it also has been proven to give invaluable insight into the workings of anomaly-induced transport \cite{Erdmenger:2008rm, Banerjee:2008th, Landsteiner:2011iq}. It also allows one to study the out-of-equilibrium evolution of strongly coupled quantum systems by means of numerical relativity in asymptotically anti-de Sitter spaces \cite{Chesler:2013lia}. 

One of the simplest but very interesting time-dependent gravity solutions is Vaidya metrics. Vaidya metrics are generated by in-falling incoherent null dust. They have very simple energy-momentum and charge distributions which allow analytic solutions of Einstein equations even in the presence of a cosmological constant. This will be our starting point. We use asymptotically anti-de Sitter charged Vaidya metrics to simulate out-of-equilibrium evolution.  We add a small magnetic field and calculate the response in the vector, axial and energy currents due to the axial anomaly in linear response. 

The response in the energy current is bound to be trivial due to the fact that it is also the conserved momentum density.  Since no additional momentum is injected into the system the value of the energy current does not change as long as momentum is conserved. This motivates us to introduce a new Vaidya type of solutions with momentum relaxation. Momentum relaxation can be introduced in a standard way by breaking translation invariance with linear spatial profiles of massless scalar fields \citep{Andrade:2013gsa}. The strength of momentum relaxation is governed by the slope of the linear scalar field background. It turns out that Vaidya-type metrics are still solutions to the equations including the energy-momentum tensor of the massless scalar fields. This has been noted before in the case of four-dimensional asymptotically AdS Vaidya metrics in Refs. \cite{Withers:2016lft,Bagrov:2017tqn}. Physically this corresponds to a sort of homogeneous distribution of heavy impurities on which momentum is destroyed. 

While we emphasize that our setup represents far-from-equilibrium physics it is still useful to compare to hydrodynamics to gain some intuition. In general transport can have some convective component due to the overall flow of the fluid. This flow is, however, impeded by the impurities (the linear scalar fields) and thus has to vanish when the external perturbations have vanished and the system approaches a new equilibrium. The response in the energy-momentum tensor and current is given as the sum of convective and anomaly induced parts. With momentum relaxation the convective part has to vanish and what is left over is only the anomaly-induced part. The anomaly-induced part is dissipation free and thus cannot be affected by impurities. In holography it has already been shown in Ref. \cite{Copetti:2017ywz} that the (equilibrium) values of anomaly induced transport coefficients are insensitive to momentum relaxation. In an effective hydrodynamics setup this
has also been shown in Ref.\cite{Stephanov:2015roa}. In the present paper this is generalized to holographic far-from-equilibrium evolution.

In holography out-of-equilibrium anomaly-induced transport has been studied before in Refs. \cite{Lin:2013sga, Ammon:2016fru, Landsteiner:2017lwm}. Let us briefly compare these to our approach. In Ref. \cite{Lin:2013sga} the authors studied a free-falling charged shell of matter in Anti de-Sitter space. While this is similar to the Vaidya approach they computed the CME response
in a quasistatic approximation whereas we directly solve for the time dependence. Reference \cite{Ammon:2016fru} studied the effects of parallel magnetic and electric fields to simulate the negative magnetoresistivity out of equilibrium. Finally Ref. \cite{Landsteiner:2017lwm} focused on the out-of-equilibrium behavior of the effects  induced by the gravitational anomaly.
We focus on directly computing the time evolution of the response in the Vaidya background by solving partial differential equations and introduce momentum relaxation as an essentially new ingredient. 

The article is organized as follows. In section \ref{sec:model} we introduce the holographic model. In section \ref{sec:background} we present AdS Vaidya-type solutions with momentum relaxation.  The linear response equations are set up in section \ref{sec:response}.  In section \ref{sec:hydro} we do a preliminary analysis based on hydrodynamics which serves to establish some intuition on the out-of-equilibrium evolution. The core of the article is section \ref{sec:results}, in which we compute the anomalous response far from equilibrium. We concentrate on two situations.  The first one is the response in the vector current in the background of axial charge and a vector magnetic field.  This is the actual CME response. The second one is the response in the axial and energy currents due to axial charge and axial magnetic field. The remaining two cases with vector charge (the chiral separation effect) turn out to be isomorphic to the previous ones. 
We analyze the relevant quasinormal mode spectrum in section \ref{sec:qnms} and compare to the direct numerical solutions of the time evolution. In section \ref{sec:discussion} we present our conclusions and outlook for further studies. Some details on the numerical integration method are supplemented in the appendix \ref{sec:appendix}.

\section{The model}\label{sec:model}

We will use a holographic model in five dimensions that includes two copies of the Maxwell action supplemented with a Chern-Simons term and three free scalar fields. The electromagnetic symmetry is represented by the gauge field $V_\mu$, the field strength of which is $F = dV$, while the axial $U(1)$ symmetry is represented by the gauge field $A_\mu$, the field strength of which is $F_5 = dA$. The Chern-Simons term is appropriately chosen to make only one of the two currents anomalous. In particular, the axial current will suffer from both a $U(1)_A^3$ and a $U(1)_A U(1)_V^2$ anomaly. Finally, the three scalars $X^I$, with $I=1,2,3$, will be later chosen to break translation symmetry and create an effective mass for the graviton. 

The action is
\begin{align}
 S =& \frac{1}{2 \kappa^2} \int d^5 x \sqrt{-g} \bigg[ R + \frac{12}{L^2} - \frac{1}{2} \partial_\mu X^I \partial^\mu X^I \nonumber\\ 
&- \frac{1}{4} F^2 - \frac{1}{4} F_5^2
+ \frac{\alpha}{3} \epsilon^{\mu\nu\rho\sigma\tau} A_\mu \left( 3 F_{\nu\rho} F_{\sigma\tau} + F_{\nu\rho}^5 F_{\sigma\tau}^5 \right) \bigg] \nonumber\\
& + S_{GH} + S_{nf} \, , 
\end{align}
where $\kappa^2$ is the Newton constant, $L$ is the AdS radius, and $\alpha$ is the Chern-Simons coupling constant, and we assume summation over repeated indices, both Lorentz and internal ones. $S_{GH}$ stands for the usual Gibbons-Hawking term
\begin{equation}
 S_{GH} = \frac{1}{\kappa^2} \int_\partial d^4 x \sqrt{-\gamma} K \, ,
\end{equation}
where the integral is over the boundary of the manifold and $\gamma$ and $K$ are the determinant of the induced metric and the trace of the extrinsic curvature, respectively. This term is included in order to make the variational problem well defined. $S_{nf}$ stands for a null-fluid action that we add by hand in order to obtain a Vaidya-like solution for the equations of motion. In the next section we discuss the details of the background construction. From now on, we will set $L=1$.

The equations of motion for the model are
\begin{align}
2 \kappa^2 Y_{(nf)}^I =& \frac{1}{\sqrt{-g}} \partial_\mu\left( \sqrt{-g} \, \partial^\mu X^I \right) \, , \\
2 \kappa^2 J_{(nf)}^\mu =& \nabla_\nu F^{\nu\mu} + 2 \alpha \epsilon^{\mu\nu\rho\sigma\tau} F_{\nu\rho} F_{\sigma\tau}^5 \, , \\
2 \kappa^2 J_{5(nf)}^\mu =& \nabla_\nu F_5^{\nu\mu} + \alpha \epsilon^{\mu\nu\rho\sigma\tau} \left( F_{\nu\rho} F_{\sigma\tau} + F_{\nu\rho}^5 F_{\sigma\tau}^5 \right) \, , \\
\kappa^2 T_{\mu\nu}^{(nf)} =& G_{\mu\nu} - \frac{6}{L^2} g_{\mu\nu} - \frac{1}{2} \partial_\mu X^I \partial_\nu X^I + \frac{1}{4} \partial_\rho X^I \partial^\rho X^I g_{\mu\nu} \nonumber\\
&- \frac{1}{2} F_{\mu\rho} {F_\nu}^\rho + \frac{1}{8} F^2 g_{\mu\nu} - \frac{1}{2} F_{\mu\rho}^5 {F_\nu^5}^\rho + \frac{1}{8} F_5^2 g_{\mu\nu} \, ,
\end{align}
where the sources in the left-hand side of the equations stand for the variation of the null-fluid action with respect to the scalar, the two different gauge fields and the metric, respectively.

From the quantum field theory point of view, the definitions of the scalar operators, consistent currents and energy-momentum tensor are 
\begin{align}
Y^I =& \lim_{\rho\to\infty}\sqrt{-\gamma} \partial_\rho X^I \,,\\
 J^i =& \lim_{\rho\to\infty} \sqrt{-\gamma} \left[ F^{i \rho} + 4 \alpha \epsilon^{ijkl} A_j F_{kl} \right] \, , \label{currents:vector}\\
 J_5^i =& \lim_{\rho\to\infty} \sqrt{-\gamma} \left[ F_5^{i \rho} + \frac{4 \alpha}{3} \epsilon^{ijkl} A_j F_{kl}^5  \right] \, , \label{currents:axial} \\
 T^{ij} = & \lim_{\rho\to\infty} 2\, \sqrt{-\gamma} \left[ - K^{ij} + K \gamma^{ij} \right] \, , \label{currents:tensor}
\end{align}
where we are implicitly using the Fefferman-Graham coordinates $ds^2 = d\rho^2 + \gamma_{ij} dx^i dx^j$.  These currents satisfy the following holographic Ward identities
\begin{align}
 \partial_i J^i &=0 \, , \label{wardid:vector} \\
 \partial_i J_5^i &= - \alpha \sqrt{-\gamma} \epsilon^{ijkl} \left( F_{ij} F_{kl} + \frac{1}{3} F_{ij}^5 F_{kl}^5 \right) \, , \label{wardid:axial} \\
 \partial_j {T^{ji}} &= F^{ij} J_j + F_5^{ij} J_j^5 - Y^I \partial^i X_I - A^i \partial_j J_5^j\, . \label{wardid:tensor}
\end{align}

\section{Background construction}\label{sec:background}

Our motivation for this work is studying the behavior out of equilibrium of anomalous transport with a small magnetic field and momentum relaxation. Therefore, our background shall represent a time-evolving homogeneous and isotropic charged state in a theory that breaks translation symmetry. The magnetic field will be later included as a perturbation on top of this background.

The simplest setup that serves our purpose consists of a black brane with time-dependent blackening factor, which in Eddington-Finkelstein coordinates has the form
\begin{equation}
 ds^2 = - f(v,r) dv^2 + 2 dv dr + r^2 d\vec{x}^2 \, ,
\end{equation}
and a linear spatial profile for the scalars, associating each one to a certain boundary coordinate
\begin{equation}
 X^1 = k x \, , \hspace{1cm}
 X^2 = k y \, , \hspace{1cm}
 X^3 = k z \, .
\end{equation}
As the scalars couple only through derivatives, the field equations and solutions are still formally translation invariant. 

We also want our background to be charged, in order to have nonzero chemical potential. As there are no magnetic or electric fields in the background, and we choose the gauge to be $V_r = A_r = 0$, the Chern-Simons terms in the equations of motion will have no contribution at the level of the background. Since these are the terms that mix both gauge fields, we can thus focus on only one of the gauge fields for the construction of the background. We stick to the vector field strength $F$ to avoid unnecessary cluttering of the notation, but the exact same discussion would apply to the axial gauge field sector.

If we impose that the expectation value of the charge density $\langle J^v \rangle$ is equal to the charge $q$, we can fix the field strength to be
\begin{equation}
 F_{rv} = \frac{q}{r^3} \, .
\end{equation}
This field strength corresponds to an external source given by
\begin{equation}
 2 \kappa^2 J_{(ext)}^v = \frac{\dot{q}}{r^3} \, ,
\end{equation}
where the dot stands for derivatives with respect to $v$.

The blackening factor $f$ can be obtained as solution of two different time-independent differential equations that arise as components of the Einstein's equations. One of them is second order and the other one is first order, but compatibility of both solutions imposes the extra integration constant from the second order one to vanish. We fix the other integration constant by comparing it to the standard form of the mass term in uncharged Vaidya metrics, finally giving
\begin{equation}
 f(v,r) = r^2 \left( 1 - \frac{k^2}{4 r^2} - \frac{2 m(v)}{r^4} + \frac{q(v)^2}{12 r^6} \right) \, .
\end{equation}
This corresponds to an external source given by
\begin{equation}
 \kappa^2 T_{vv}^{(ext)} = \frac{3 \dot{m}}{r^3} - \frac{q \dot{q}}{4 r^5} \, .
\end{equation}
One could wonder why the mass and charge are allowed to vary with time but the momentum relaxation coefficient is not. It is important to note in this regard that their origins are very different. As it was discussed in the first paragraph of this section, momentum relaxation is a feature of the theory, while the charge and the mass, necessary to have a black brane, are properties of the state. In fact, their appearance in the context of holography is very different. While the momentum relaxation coefficient is fixed when sourcing the scalars, the mass and the charge appear as integration constants in the solution of the differential equations.

The thermodynamics of the system are defined by the chemical potential and the temperature, which we compute as the difference between the boundary value and the horizon value of the zeroth component of the gauge field and the Hawking temperature of the black brane, respectively. Their values are
\begin{equation}
 \mu = \frac{q}{2 r_H^2} \, , \hspace{10pt} T = \frac{1}{4 \pi} \left( \frac{k^2}{2 r_H} + \frac{8 m}{r_H^3} - \frac{q^2}{2 r_H^5} \right) \, , \label{muT}
\end{equation}
where $r_H$ stands for the position of the apparent horizon\footnote{We do not consider the event horizon since its location could be changed by events that happen long after the end of our numerical simulations.}. Please note again that all this discussion would be the same in the presence of axial charge exchanging $q$ by $q_5$, or adding both contributions if both charges were present.

One obvious concern may arise. While chemical potential and temperature cannot be defined in out-of-equilibrium setups, definitions \eqref{muT} in this Vaidya background are valid at all times and react instantaneously to the changes in mass and charge. We would like to make two comments on this. On one hand, although one could feel tempted to take these expressions as true also out-of-equilibrium, it is important to understand that they are ill defined in such regimes and we will only use them as a near-equilibrium approximation to detect true out-of-equilibrium behavior. On the other hand, we want to look at the out-of-equilibrium behavior of anomalous transport. We use this Vaidya background in order to simplify the computations. The fact that the background seems to react to the changes instantaneously does not alter the fact that the computed one-point functions present clear out-of-equilibrium behavior. We would not expect notable deviations in the qualitative picture if the procedure involved backgrounds that are generated by numerically solving for collapse-type space-times as it was done in Ref.\cite{Landsteiner:2017lwm}.

From now on, we will change our radial coordinate to be
\begin{equation}
 u = \frac{1}{r} \, ,
\end{equation}
and we denote the derivative with respect to this new coordinate by a prime. We use this coordinate in order to simplify the numerical procedure.

\section{Linear response computations}\label{sec:response}

We are going to probe the out-of-equilibrium behavior of the system by switching on a small constant magnetic field, either vector or axial, and the minimal set of fluctuations required by consistency of the equations of motion at linear order in the magnetic field. To carefully perform the perturbative computation, we include an infinitesimal coefficient $\epsilon$ of which the powers will account for the order in the expansion and we will drop all powers from the second one on.

Without loss of generality, we set our magnetic fields along the $z$ axis, such that $F_{xy} = \epsilon B$ or $F_{xy}^5 = \epsilon B_5$ in each case. The fluctuations  that will be switched on are $V_z = \epsilon V$, $A_z = \epsilon A$, $g_{vz} = \epsilon h/u^2$, and $X^3 = kz + \epsilon Z$, grouped in different subsets depending on the particular case, as will become evident below.

It is straightforward to check that the four different cases with vector or axial charge and vector or axial magnetic field reduce to two different sets of equations. The first set exists for the two cases with vector magnetic field and the second set appears for the two cases with axial magnetic field, and they read
\begin{align}
B: 0 &= dV' - \frac{1}{2 u} dV + \frac{u f}{4} V' - 4 \alpha B q_5 u^2 \, , \label{Veqn} \\
B_5:  0 &= dA' - \frac{1}{2 u} dA + \frac{u f}{4} A' - 4 \alpha B_5 q_5 u^2 + \frac{q_5 u}{2} H \, , \label{Aeqn} \\
0 &= dh' + \frac{5 u f + u^2 f'}{2} H - q_5 u^3 dA - k dZ + k^2 h \, , \label{heqn} \\
0 &= dZ' - \frac{3}{2u} dZ + \frac{3u f}{4} Z' + \frac{3 k}{2 u} h - \frac{k}{2} H \, , \label{Zeqn} \\
0 &= \left( \frac{H}{u^3} \right)' - q_5 A' + \frac{k}{u^3} Z' \, , \label{cstr}
\end{align}
where $H = h'$ and $d$ stands for directional derivatives along the outgoing null geodesics
\begin{equation}
 d = \partial_v - \frac{u^2 f}{2} \partial_u \, .
\end{equation}
The other equations in each of the sets of equations are obtained by $q5\leftrightarrow q$ and $A\leftrightarrow V$. More on this will be discussed below. 

Equation \eqref{cstr} involves no time derivative and thus it can be understood as a constraint that has to be fulfilled at all times. This form of \eqref{heqn} is obtained after using the constraint to exchange time derivatives by the new operator $d$. It is straightforward to see that the momentum relaxation coefficient acts simultaneously as a coupling between the scalar and the metric perturbation and a mass for the metric. If there were no momentum relaxation, the scalar would decouple from the system of equations, as it will also become obvious from the quasinormal modes. Please notice that, although not explicitly, momentum relaxation is also relevant in the equations of the gauge fields through the blackening factors.

It is mandatory to check the compatibility of the equations of motion and the constraint for the case with axial magnetic field. It can be seen that
\begin{align}
- \dot{q_5} A' =& d [\mathrm{Eq.}\ref{cstr}] - \partial_u \left( \frac{[\mathrm{Eq.}\ref{heqn}]}{u^3} \right) \nonumber\\
&- \partial_u \left( \frac{u^2 f}{2} \right) [\mathrm{Eq.}\ref{cstr}] - \frac{2 k}{u^3} [\mathrm{Eq.}\ref{Zeqn}] \, . 
\end{align}
Therefore, the constraint and the equations of motion are compatible only in the case in which the axial charge does not vary with time. This is not a feature of momentum relaxation, since it is true even for the momentum-preserving case, $k=0$.

The definition of the directional derivatives along the outgoing geodesics gives us a time-evolution equation for the different perturbations by solving for the $v$ derivative
of each field. The use of this directional derivative is according to the well-known method of characteristics, in which one uses information about the trajectories along which the information of the solution is transported to simplify the form of the equations to be solved. In the case with $B_5$, the constraint could, in principle, be used as the equation for $h$, but we got better accuracy by using the same procedure for $A$, $h$, $Z$ and then the constraint can be used as a test that the time evolution was correct.

So far, we have only discussed two of the four possible situations. The case with $q$ and $B$ can be obtained from the case with $q_5$ and $B$ by exchanging $V$ by $A$ and $q_5$ by $q$ in \eqref{Veqn}. The case with $q$ and $B_5$ can be obtained by exchanging $A$ by $V$ and $q_5$ by $q$ in \eqref{Aeqn}-\eqref{cstr}. Thus, the-out-of-equilibrium behavior is grouped according to the magnetic field being vector or axial and we only need to make the computations for the two cases considered above. 

However, this is rather peculiar because, according to the type of anomaly, the classification is different: the case with $q_5$ and $B_5$ stems from the $U(1)_A^3$ while the other three come from the $U(1)_A U(1)_V^2$. We can better understand this by looking at the expressions these effects have in-equilibrium in terms of the anomaly coefficients
\begin{align}
 \vec{J}_a &= d_{abc} \frac{\mu_b}{4 \pi^2} \vec{B}_c \, , \\
 \vec{J}_\epsilon &= d_{abc} \frac{\mu_a \mu_b}{8 \pi^2} \vec{B}_c  \, ,
\end{align}
where $a,b,c=V,5$ and $\vec{J}_\epsilon$ is the energy current. It can be seen, for example, for a single Dirac fermion ($d_{VV5} = d_{V5V} = d_{5VV} = d_{555} = 2$) that, while a vector magnetic field never produces an energy current unless both chemical potentials are nonzero, the axial magnetic field produces it as long as one of the chemical potentials is present.

We need to solve the two different systems numerically. To do that, we need boundary conditions, which we obtain from the asymptotic solutions near the boundary of AdS, and also initial conditions, which we obtain from the fact that the system is in equilibrium at the beginning. All the details about the numerical computations are included in Appendix \ref{sec:appendix}.

\section{Hydrodynamic considerations}\label{sec:hydro}

To detect true out-of-equilibrium phenomena, we can compare our results to near-equilibrium ones obtained from hydrodynamics. We start by writing down the constitutive relations for the different currents,
\begin{align}
 J^\mu =& \rho u^\mu + \sigma B^\mu + \sigma_5 B_5^\mu \, , \label{eq:Jhydro}\\
 J_5^\mu =& \rho_5 u^\mu + \tilde{\sigma} B^\mu + \tilde{\sigma}_5 B_5^\mu \, , \label{eq:J5hydro}\\
 T^{\mu\nu} =& ( \epsilon + p ) u^\mu u^\nu + p \eta^{\mu\nu} + \xi u^{(\mu} B^{\nu)} + \xi_5 u^{(\mu} B_5^{\nu)} \, ,
 \label{eq:Thydro}
\end{align}
where parentheses on the indices stand for symmetrization $A_{(\mu} B_{\nu)} = A_\mu B_\nu + A_\nu B_\mu$. Please note that the standard symmetrization also involves a combinatorial factor in the denominator, but we do not include it because our purpose is just to obtain more compact expressions.

Typically, relativistic hydrodynamics has an ambiguity on the choice of frame, which in holography presents itself as a choice of boundary conditions. In the cases with axial magnetic field and momentum relaxing parameter $k$ different from zero, regularity conditions at the horizon impose one more condition on the independent modes of the series expansion: the metric perturbation has to be zero at the horizon. Therefore, it can be said that momentum relaxation chooses a preferred frame and that this frame is the disorder rest frame (the ``no-drag'' frame of Ref.\cite{Stephanov:2015roa}). For convenience, we also impose this condition in the rest of the cases, in which we have freedom to choose the frame.

Once we choose the frame, we stick to it for the whole computation. In and near equilibrium, the transport coefficients in this frame are given by 
\begin{alignat}{5}
 \sigma &=&& \, 8 \alpha \mu_5 \, , \hspace{.5cm} &&\sigma_5 &&=&& \, 8 \alpha \mu \, , \label{coeff:sigma} \\
 \tilde{\sigma} &=&& \, 8 \alpha \mu \, , \, \hspace{.5cm} &&\tilde{\sigma}_5 &&=&& \, 8 \alpha \mu_5 \, , \label{coeff:tildesigma} \\
 \xi &=&& \, 4 \alpha \mu \mu_5 \, , \, \hspace{.5cm} &&\xi_5 &&=&& \, 4 \alpha \mu_5^2 \, . \label{coeff:xi}
\end{alignat}
Any deviation in the one-point functions of the quantum currents \eqref{currents:vector}-\eqref{currents:tensor} from these expressions will be taken as a sign of out-of-equilibrium physics. The details of the numerical computation of those expectation values are also reported in appendix \ref{sec:appendix}.

In the cases with axial magnetic field, we will be looking at both the axial current and the energy current. In relativistic hydrodynamics, the energy current $T^{0i}$ is equal to the momentum density $P^i = T^{i0}$. The latter is usually a conserved quantity, except in the case with momentum relaxation.  

Without momentum relaxation, the response in the energy current is bound to be trivial. The chemical potentials change, of course, but the different anomaly-induced energy current is compensated by a convective part due to flow in \eqref{eq:Thydro}.
This convective component is subject to dissipation in the case with momentum relaxation. 
We cannot compute the exact value of the fluid velocity, because we only have access to the one-point functions of the different currents but not to the anomalous and convective contributions separately.

However, we can compute what the flow would be near equilibrium from the point of view of hydrodynamics  with momentum conservation. Since the one-point function of the $0i$-components of the energy-momentum tensor does not change, it always has its initial value. This means, according to the constitutive relation in \eqref{eq:Thydro}, that
\begin{equation}
 \langle T^{0z} \rangle =  4 \alpha (\mu_5^{in})^2 B_5  = ( \epsilon + p ) v_z + 4 \alpha \mu_5^2 B_5 \, ,
\end{equation}
where the superscript $in$ means it is the initial value of the axial chemical potential. We can solve for the fluid velocity 
\begin{equation}
 v_z = \frac{4 \alpha B_5}{\epsilon + p} \left[ (\mu_5^{in})^2 - \mu_5^2 \right] \, .
\end{equation}
Once we have computed the fluid velocity, we can use the constitutive relation of the axial current \eqref{eq:J5hydro} to see what the value of the one-point function will be according to hydrodynamics. It gives
\begin{equation}
 \langle J_5^z \rangle = \rho_5 v_z + 8 \alpha \mu_5 B_5 = 8 \alpha \mu_5 B_5 + \frac{4 \alpha B_5 \rho_5}{\epsilon + p} \left[ (\mu_5^{in})^2 - \mu_5^2 \right] \, .
\end{equation}
At this point, we use again the constitutive relations and the holographic dictionary with the background metric and gauge field to obtain the transport coefficients
\begin{align}
 \epsilon &= \langle T^{00} \rangle = 6 m \, , \\
 p &= \langle T^{ii} \rangle = 2 m \, , \\
 \rho_5 &= \langle J^0 \rangle = q_5 \, ,
\end{align}
and substitute to obtain the equilibrium value of the axial current considering flow with momentum conservation
\begin{equation}
 \langle J_5^z \rangle = 8 \alpha \mu_5 B_5 + \frac{\alpha B_5 q_5}{2 m} \left[ (\mu_5^{in})^2 - \mu_5^2 \right] \, .
\end{equation}
We use these hydrodynamic expressions for the currents as benchmarks for near-equilibrium evolution in the following
(see Figs. \ref{fig:B5tau}, \ref{fig:B5KJ} and \ref{fig:B5K1J}) in which $\mu_5$ is defined as in \eqref{muT}.

\section{Results}\label{sec:results}

In all the cases studied, we performed a quench in the mass of the form
\begin{equation}\label{eq:quench}
 m = m_0 + \frac{m_f - m_0}{2} \left( 1 + \tanh \left( \frac{v}{\tau} \right) \right) \, .
\end{equation}
We chose the masses in order to fix initial and final horizon positions to $u_H^\mathrm{initial} = 1.0$ and 
$u_H^\mathrm{final} = 0.8$, respectively, so the exact values depend on $q_5$ and $k$ for each run. 
All dimensionful quantities quoted from now on should be understood as expressed in units set by the value of the initial horizon $u_H^\mathrm{initial}$.

The condition that the charge had to remain constant only appeared for the case with axial magnetic field, but we fixed its value to $q_5 = 1.0$ for all the cases. The benefit of this is that all the runs have the same initial and final chemical potential and, therefore, they can be compared. However, the initial and final temperatures will not be the same for the different cases in which we compare runs with different values of $k$. The specific value of the charge has no special meaning and only affects the results by a normalization. Nevertheless, we used this value in order for the flow to have a sufficiently large value that can be clearly seen in the results.

\subsection{Momentum conservation}
The first analysis we decided to make was reminiscent of some of the results in Ref.\cite{Landsteiner:2017lwm}. We wanted to see the impact that the time span of the quench $\tau$ had on the equilibration process on both cases without momentum relaxation, by looking at the results for several different values of this parameter. 

\begin{figure}
 \includegraphics[width=0.5\textwidth]{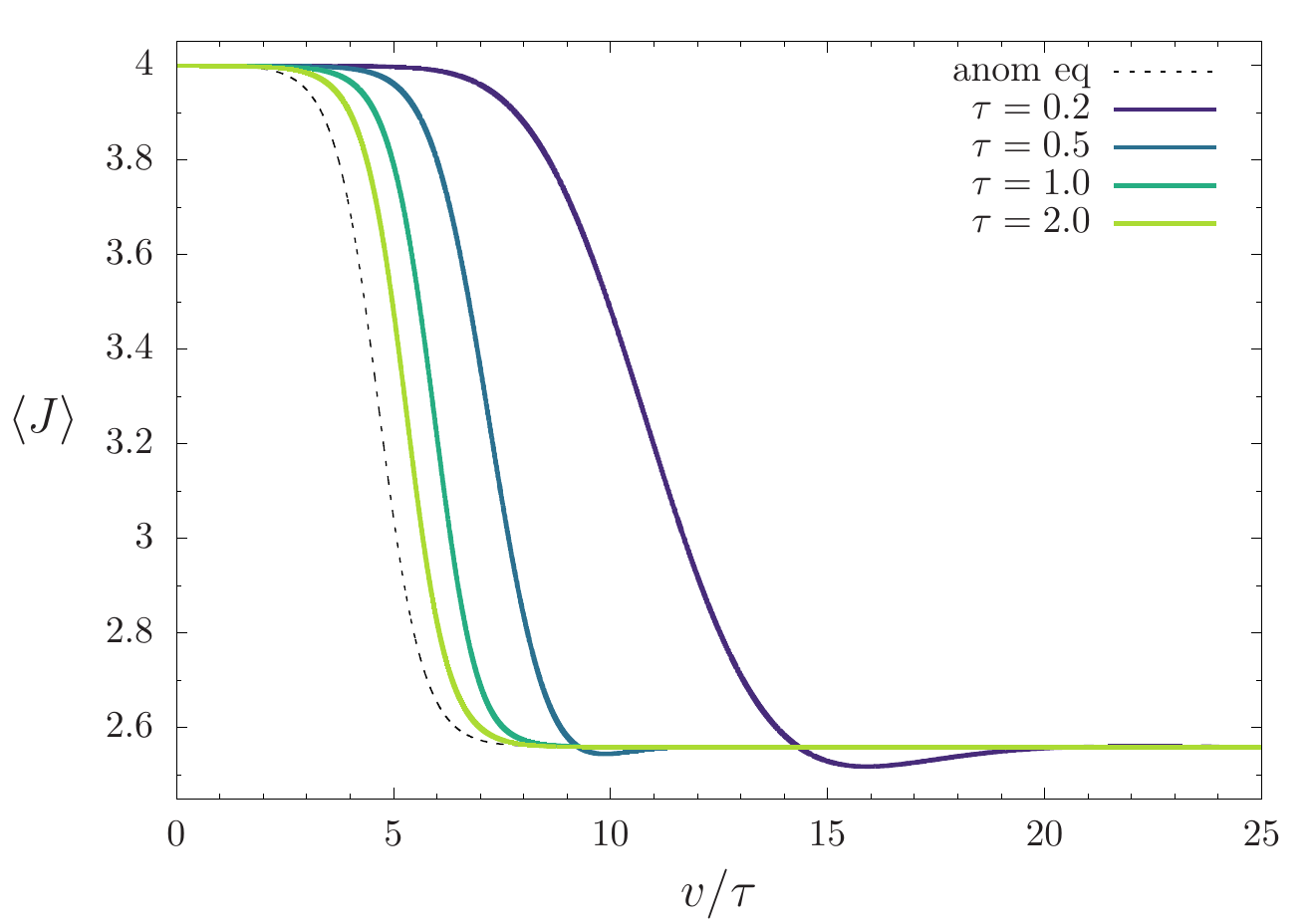}
 \caption{Out-of-equilibrium electromagnetic current for different values of the time span of the quench with no momentum relaxation. The black dashed line is the near-equilibrium approximation we use as a reference to signal out-of-equilibrium behavior.}
 \label{fig:Btau}
\end{figure}

In Fig. \ref{fig:Btau}, we show the results for the case with vector magnetic field. It can be seen that there are two different regimes. The first one is characterized by overshooting before the equilibration finishes.  We call them fast quenches.
If the quench is fast enough it also shows what we call delay.
By delay we mean that the response in the current builds up essentially after the time-dependent perturbation (quench) \eqref{eq:quench} has already finished. We will quantify this delay more precisely below.
The other regime, slow quenches, shows smooth monotonic behavior and have no delay. In the limit of very large $\tau$ it approaches the near-equilibrium approximation based on \eqref{muT}. The transition between both regimes appears at around $\tau \approx 1$. We note that this is significantly simpler than the behavior observed in Ref.\cite{Landsteiner:2017lwm} in the case of the gravitational anomaly-induced CME, in which three different regimes of fast, intermediate and slow quenches could be distinguished.\footnote{Three regimes of fast, intermediate and slow quenches have also been found in studies of thermalization of two-point functions in \cite{Banerjee:2016ray} and \cite{Cartwright:2019opv}. It would certainly be interesting to investigate in more detail the origin of the differences to these previous studies. This goes however beyond the purpose of the present work and we leave this question for future research.}

\begin{figure}
 \includegraphics[width=0.5\textwidth]{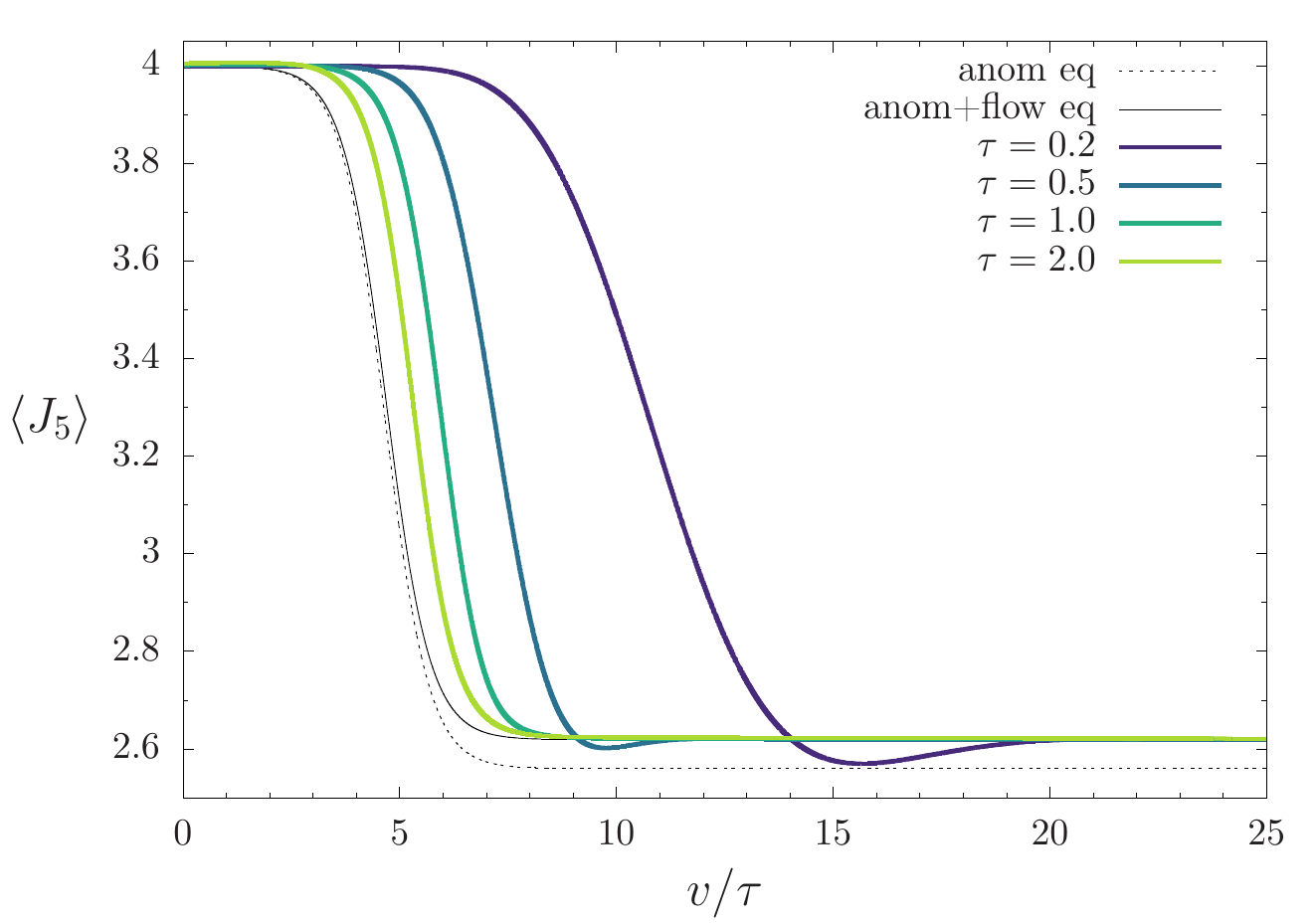}
 \caption{Out of equilibrium axial current for different values of the time span of the quench. The black dashed line is the near equilibrium approximation we use as a reference to signal out of equilibrium behavior.}
 \label{fig:B5tau}
\end{figure}

In Fig. \ref{fig:B5tau}, we show the results for the case with axial magnetic field. Both regimes can be again observed in this case and we can see that the equilibration times are essentially the same ones. However, the rest of the behavior is different due to the appearance of nonvanishing flow, as discussed above. The final equilibrium value is somewhat larger
than what could be expected from applying the CME formula alone. We attribute this to the convective flow component present
in the final state as outlined in the section \ref{sec:hydro}. The fluid velocity in the final state can be computed from Eq. (\ref{eq:Thydro}) by demanding that the final and initial energy currents (momentum density) are the same. Once this flow component is taken into account the
axial current in the final state matches the expectation from hydrodynamics perfectly (\ref{eq:J5hydro}) as can be seen from the black continuous line in Fig. \ref{fig:B5tau}.

In the case of fast quenches, the overshooting can be checked to never cross below the value the current would have if only the anomalous contribution was present. 

\subsection{Momentum relaxation}
\begin{figure}
 \includegraphics[width=0.5\textwidth]{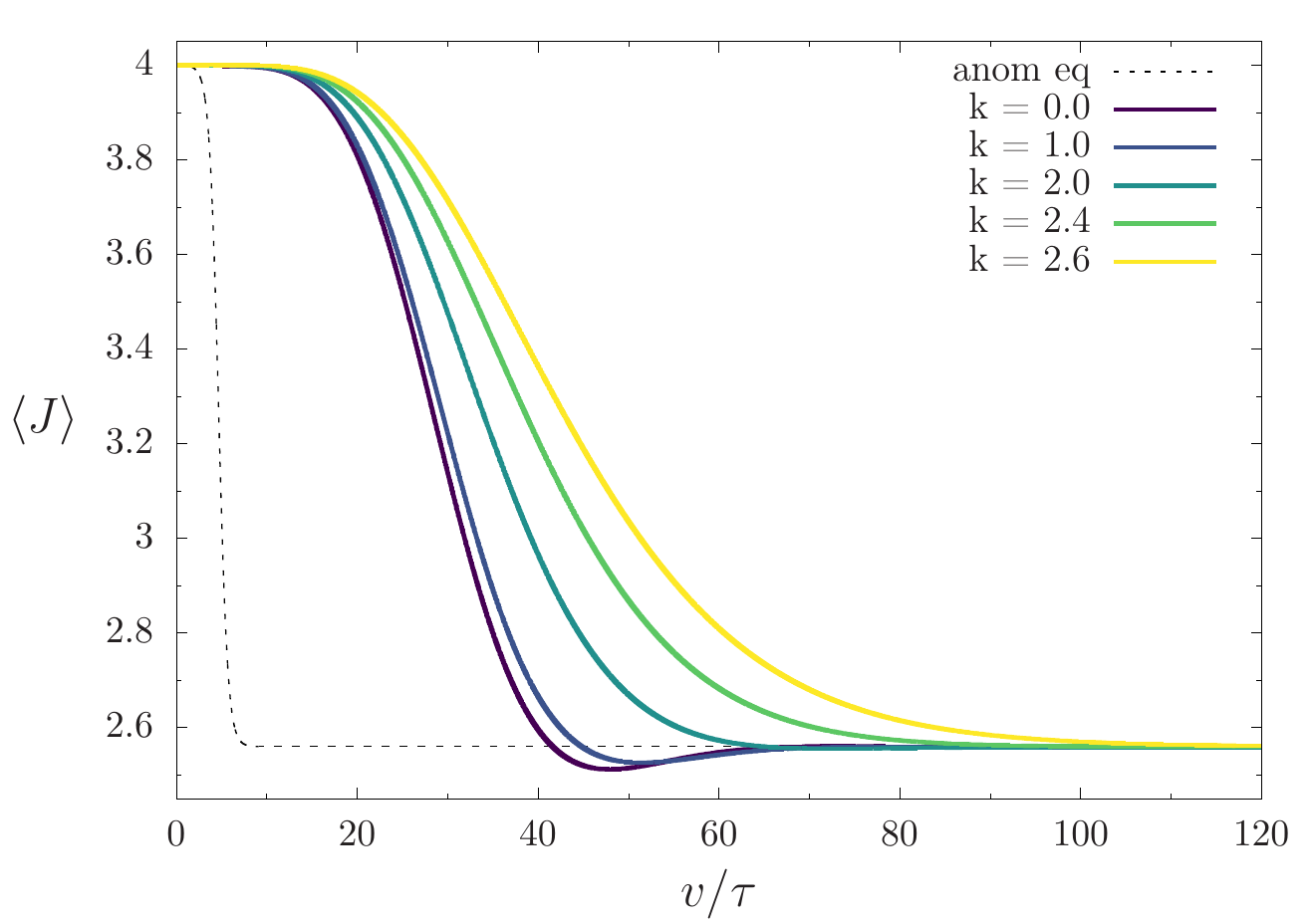}
 \caption{Out-of-equilibrium electromagnetic current for different values of the momentum relaxation parameter $k$. The black dashed line is the near-equilibrium approximation we use as a reference to signal out-of-equilibrium behavior.}
 \label{fig:BK}
\end{figure}

We now look at the results with momentum relaxation. To do that, we compare the quenches for several different values of $k$, always keeping $\tau = 0.05$. We start by looking at the case with vector magnetic field. We see again in Fig.\ref{fig:BK} the appearance of two different regimes for the current, although now the transition happens for $k \lesssim 2$. 

\begin{figure}
 \includegraphics[width=0.5\textwidth]{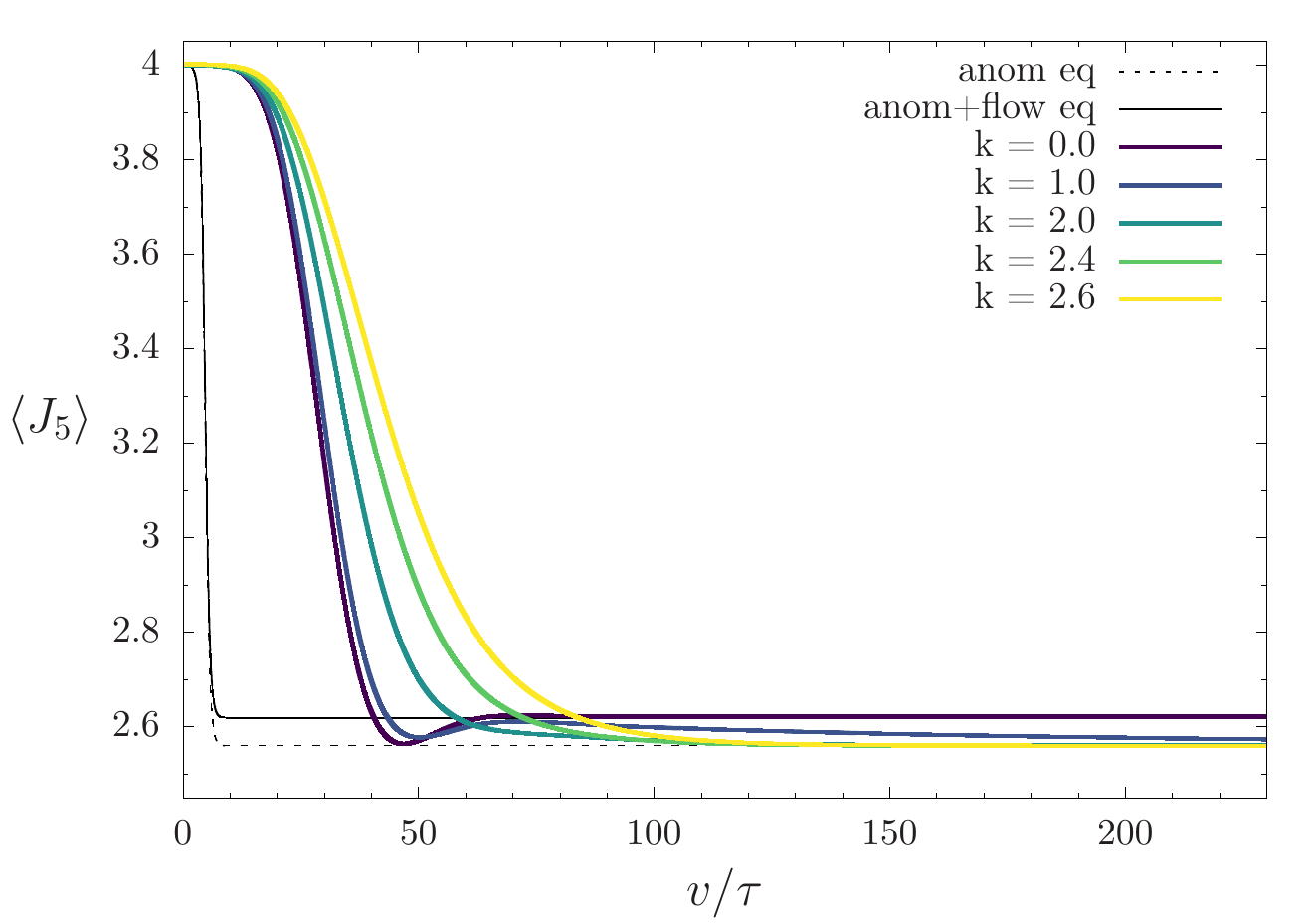}
 \caption{Out-of-equilibrium axial current for different values of the momentum relaxation parameter $k$. The black dashed line is the near-equilibrium approximation we use as a reference to signal out of equilibrium behavior.}
 \label{fig:B5KJ}
\end{figure}

\begin{figure}
 \includegraphics[width=0.5\textwidth]{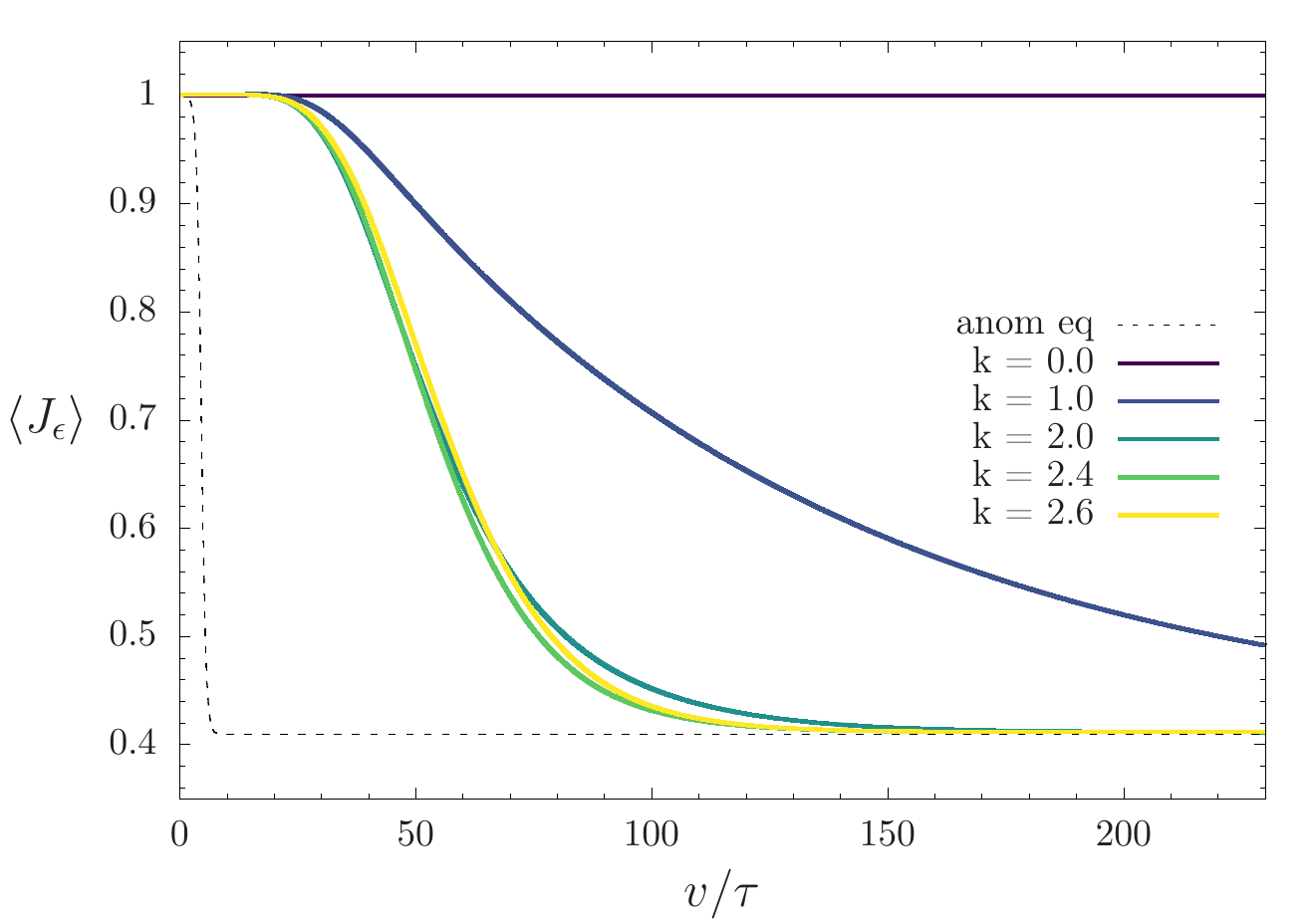}
 \caption{Out-of-equilibrium energy current for different values of the momentum relaxation parameter $k$. The black dashed line is the near equilibrium approximation we use as a reference to signal out-of-equilibrium behavior.}
 \label{fig:B5KT}
\end{figure}

For the case with axial magnetic field, we obtain Figs. \ref{fig:B5KJ} and \ref{fig:B5KT}. In the current, it can be observed that there are now two equilibration processes, with different timescales, and their interaction gives a richer structure. One of them is already present for $k=0$ and is produced by the change in the mass. The other one, only present for $k\neq0$ is precisely related to the disappearance of flow due to momentum relaxation. For small values of $k$, like $k=1.0$ in the plots, this second process is so slow that the current almost equilibrates to the value with full flow before slowly decreasing toward the result with no flow. For bigger values of $k$, though, the momentum relaxation is faster and the two processes cannot be seen as independent, although now the whole process is much longer than for the case without momentum relaxation.

The plot for the energy current shows a much simpler structure. For no momentum relaxation, the energy current stays constant as it is equal to the momentum density, which is a conserved current. For $k\neq0$ it interpolates between the initial and final equilibrium values of the anomalous contribution, and the process is faster for bigger $k$'s. However, there is a point, around $k=2.5$ where this trend changes and the process starts being slower for bigger $k$. As we
will see this is in agreement with the quasinormal mode structure of the system. 

\begin{figure}
 \includegraphics[width=0.5\textwidth]{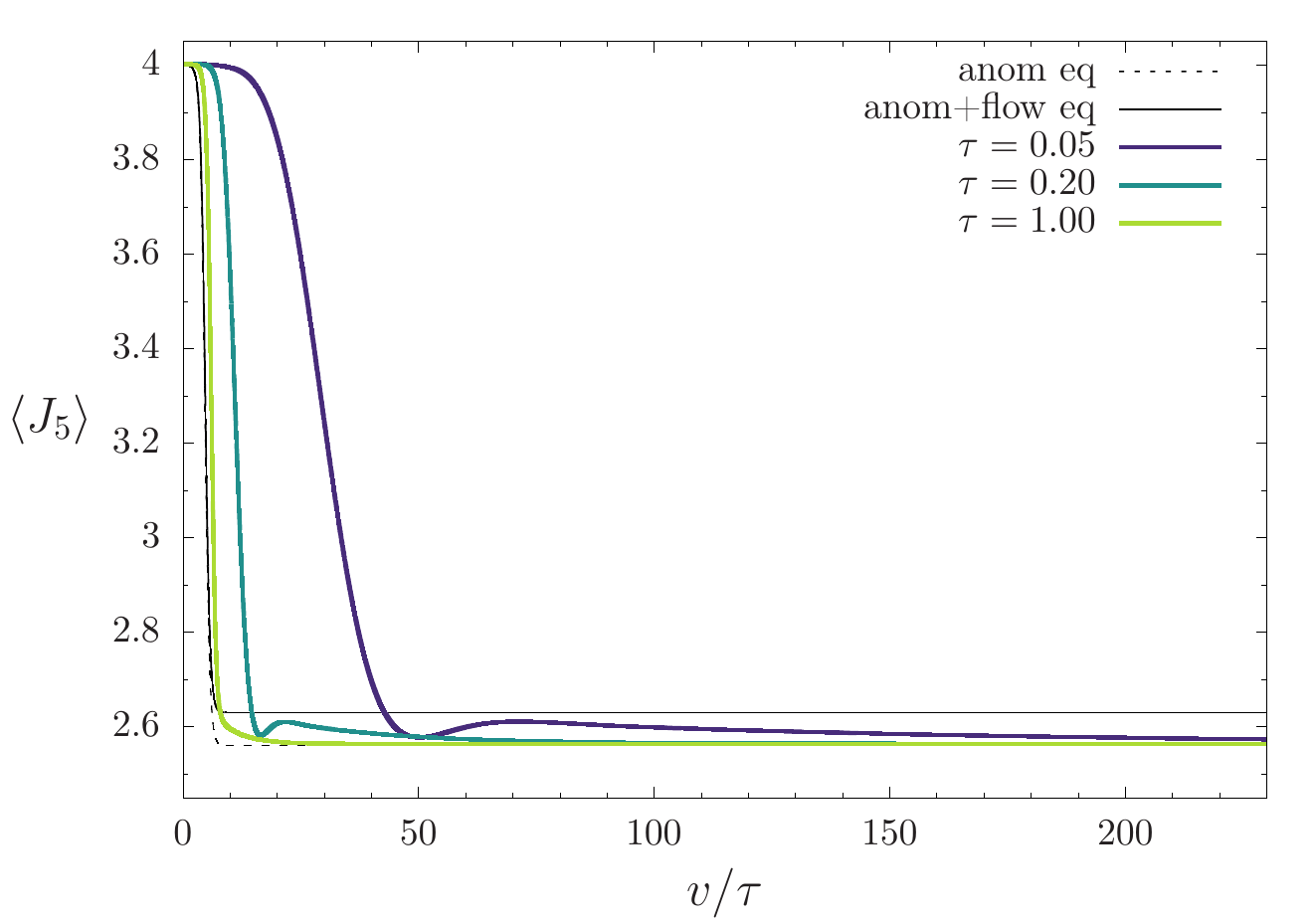}
 \caption{Out of equilibrium axial current for different values of the time span of the quench with momentum relaxation parameter $k = 1$. The black dashed line is the near equilibrium approximation we use as a reference to signal out of equilibrium behavior.}
 \label{fig:B5K1J}
\end{figure}

\begin{figure}
 \includegraphics[width=0.5\textwidth]{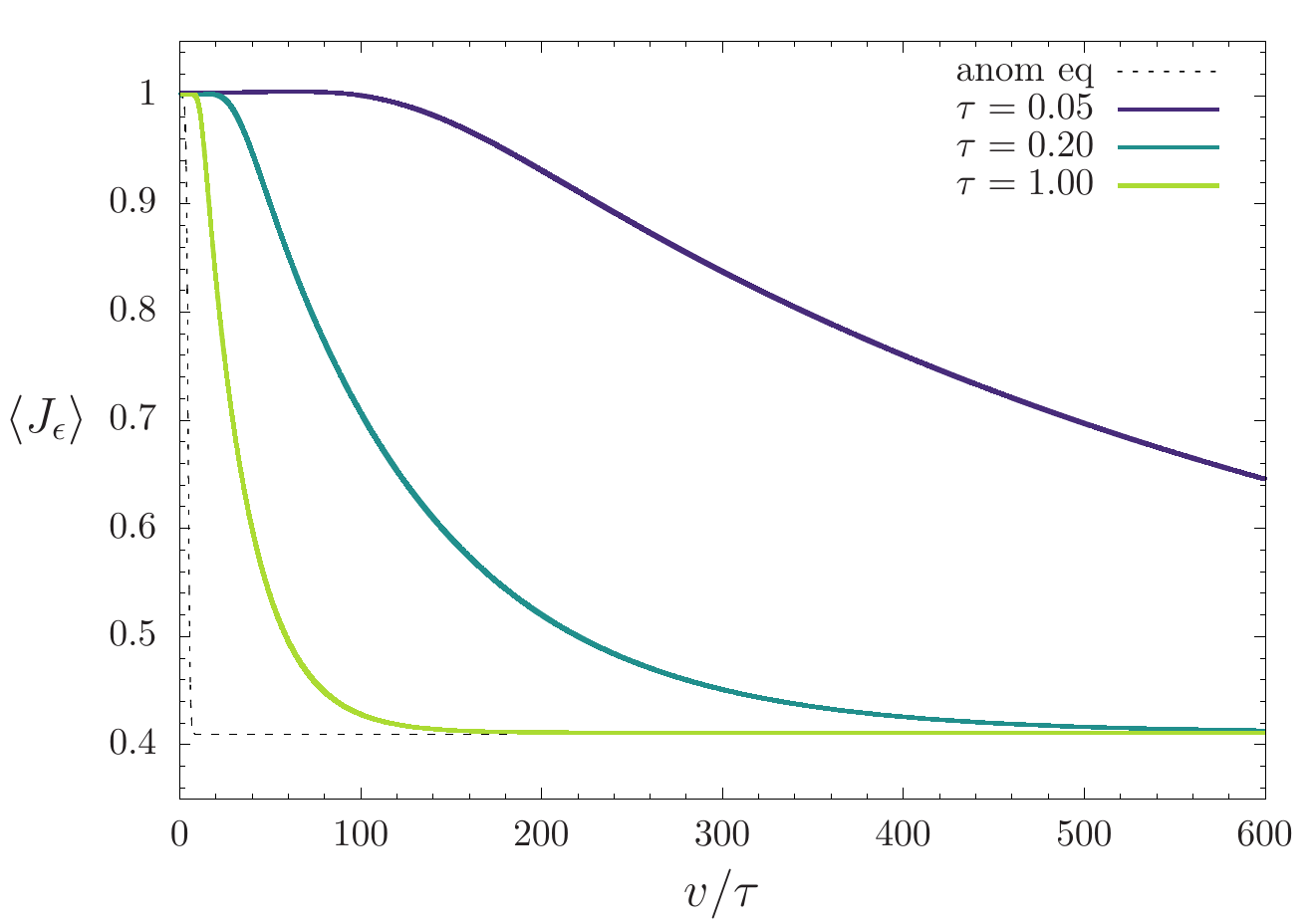}
 \caption{Out of equilibrium energy current for different values of the time span of the quench with momentum relaxation parameter $k=1$. The black dashed line is the near equilibrium approximation we use as a reference to signal out of equilibrium behavior.}
 \label{fig:B5K1T}
\end{figure}

Finally, to better understand the structure of the cases with axial magnetic field and momentum relaxation, we now plot the results for different values of $\tau$ and a fixed value of $k=1$ in  Figs. \ref{fig:B5K1J} and \ref{fig:B5K1T}. It can be seen that, as in all previous cases, the results are closer to the near-equilibrium approximation as $\tau$ becomes larger. However, the structure of the time evolution of the axial current is richer. As $\tau$ becomes smaller, it can be seen that momentum relaxation is not dominant enough to fully suppress the appearance of flow. Therefore, the response first tends to equilibrate to the value it would have if $k=0$, only to subsequently deviate from that value and end up reaching the final equilibrium value through the disappearance of flow.

\subsection{Delay}

One interesting feature in the obtained results is what we call delay. It can be defined as the time lapse after the quench has finished and before the response in the different currents starts to build up. It appears for fast quenches in both the vector and axial currents. However, we will concentrate on the vector case because it is the one that could potentially have interesting phenomenological implications for heavy ion collisions, as we will comment on below. 

In particular, we define delay in the following way. We considered the quench to be finished when the value of $8 \alpha \mu B$ with $\mu$ as in (\ref{muT}) deviates less than $0.1\%$ from the final equilibrium value. In an analogous way, we considered the buildup in the current to start when its value deviates more than $0.1\%$ from the initial value. The delay $\Delta$ is equal to the difference in $v$ between those two instants. 

\begin{figure}
 \includegraphics[width=0.5\textwidth]{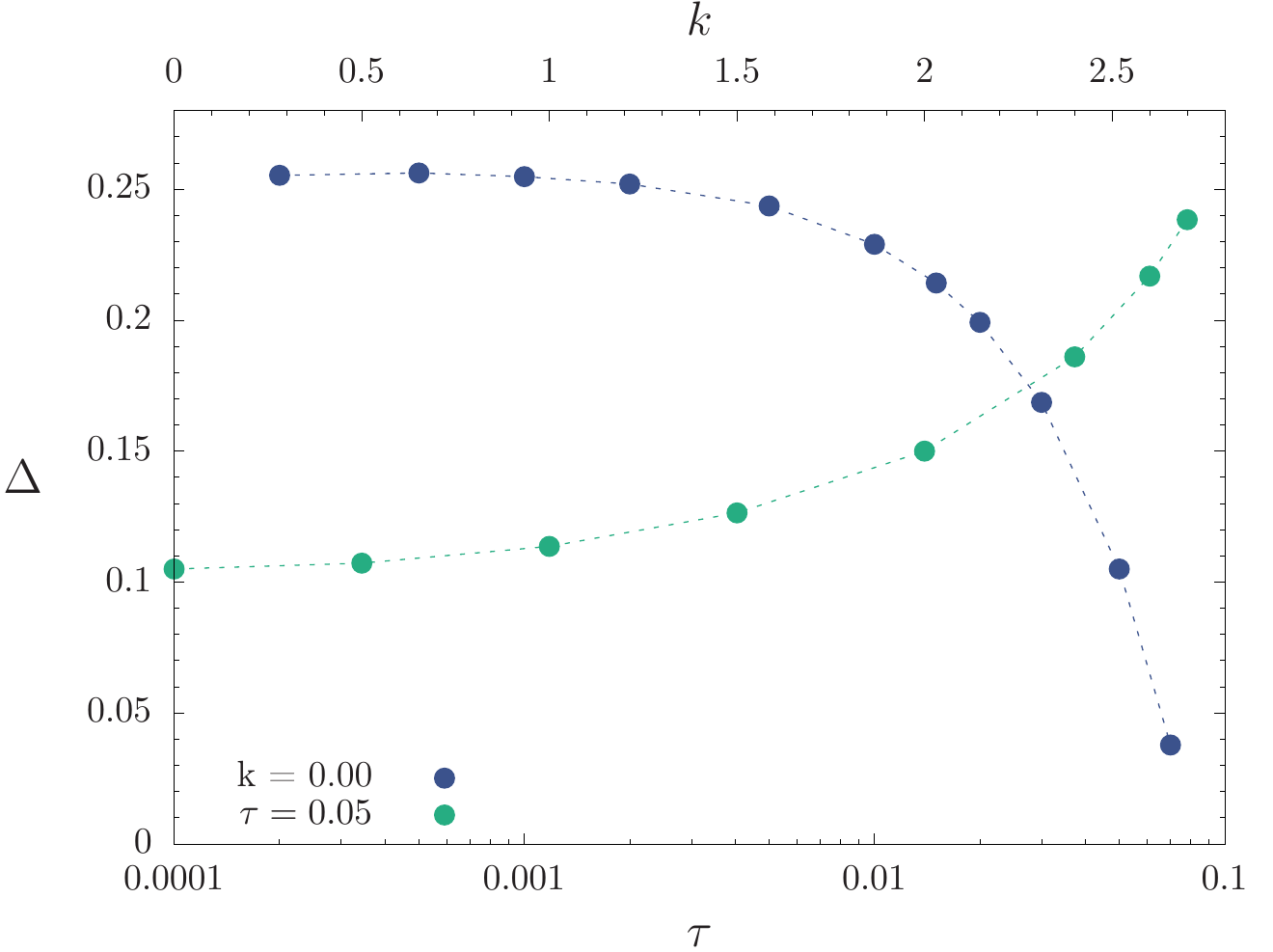}
 \caption{Delay presented as a function of the length of the quench $\tau$ for the case with $k=0$ and as a function of the momentum relaxation parameter $k$ for the case with $\tau = 0.05$.}
 \label{fig:delays}
\end{figure}

We expect the delay to depend on the momentum relaxation coefficient $k$ and the time length of the quench $\tau$. The results are presented in Fig. \ref{fig:delays}. In the first set of points, we show the dependence on $\tau$ for fixed $k = 0$. In the second set of points, we show the dependence on $k$ for fixed $\tau = 0.05$. It can be seen that the delay becomes bigger for bigger $k$. However, it becomes smaller for bigger $\tau$. In particular, it can become negative for big $\tau$ (roughly around 0.1 for the case with $k=0$), which means that there is no delay and the current starts to build up before the quench is finished. Thanks to the logarithmic scale, it can be seen that for very fast quenches at fixed $k$ the value gets to a plateau and it has a well-defined finite limit for $\tau \to 0$.

This could have interesting implications for the search of anomalous transport in heavy ion collisions. Our results suggest that the response in the current starts to build up some time after the end of the equilibration process for sufficiently fast quenches. Indeed, equilibration or hydrodynamization is supposed to be fast in heavy ion collisions. At the
same time the lifetime of the magnetic field is finite. It is also known that the lifetime of the magnetic field is
shorter for higher-energy collisions. While it is sometimes assumed that the net effect of stronger magnetic field and shorter lifetime compensate each other our results might point into a different direction.  
If the lifetime of the magnetic field
is short the delay might mean that no CME current can actually be built up before the magnetic field decays. This could,
in principle, allow the case in which the CME signal is suppressed at high energies at the LHC even if it is observable at the lower RHIC energies. Indeed, the current results from experimental searches for CME signals at RHIC and LHC allow 
precisely such an interpretation \cite{Zhao:2019hta}. 

The present model is certainly too simplistic to allow application to a more realistic situation.  Our results call, however, for further detailed studies with more phenomenological input, such as the finite lifetime of the magnetic field. 

\section{Quasinormal Modes}\label{sec:qnms}

Since the Eqs. (\ref{Veqn})-(\ref{cstr}) are linear in the fields one might expect that the time evolution can
be reasonably well described in terms of quasinormal modes. However, such an analysis is complicated due to the explicit
time dependence of the blackening factor. A complete quasinormal mode analyses would therefore have to include also the
modes stemming from the nonlinear metric equations. Instead we will study a somewhat simpler problem. We consider the final
equilibrium state and ask how fast a generic perturbation of that state decays. The equations for the perturbations are 
the same as  the Eqs. \eqref{Veqn}-\eqref{cstr} without the terms with the magnetic field.
The equations for the quasinormal modes follow then by writing
\begin{equation}
d = -i\omega - \frac{u^2 f}{2} \partial_u\,.
\end{equation}
The resulting (system of) equations are solved imposing regularity of the solutions on the horizon and vanishing non-normalizable mode on the boundary. In the case of the system of equations (\ref{Aeqn})-(\ref{cstr}) one needs to
construct three linearly independent solutions. The constraint allows, however, only for two independent solutions. 
A trivial third solution can be found by choosing $h = -i \omega$ and $Z=k$. 

The proper boundary conditions on the horizon for the gauge fields and the scalar field are that they take finite but nonvanishing values whereas the metric perturbation $h$ has to vanish on the horizon. With these boundary conditions the solutions can be found by numerical integration. In the case of the coupled system of equations the quasinormal modes are found by setting the determinant of the matrix spanned by the three linearly independent solutions to zero at the boundary \cite{Amado:2009ts,Kaminski:2009dh}. 

\begin{figure}
 \begin{center}
 \includegraphics[width=0.5\textwidth]{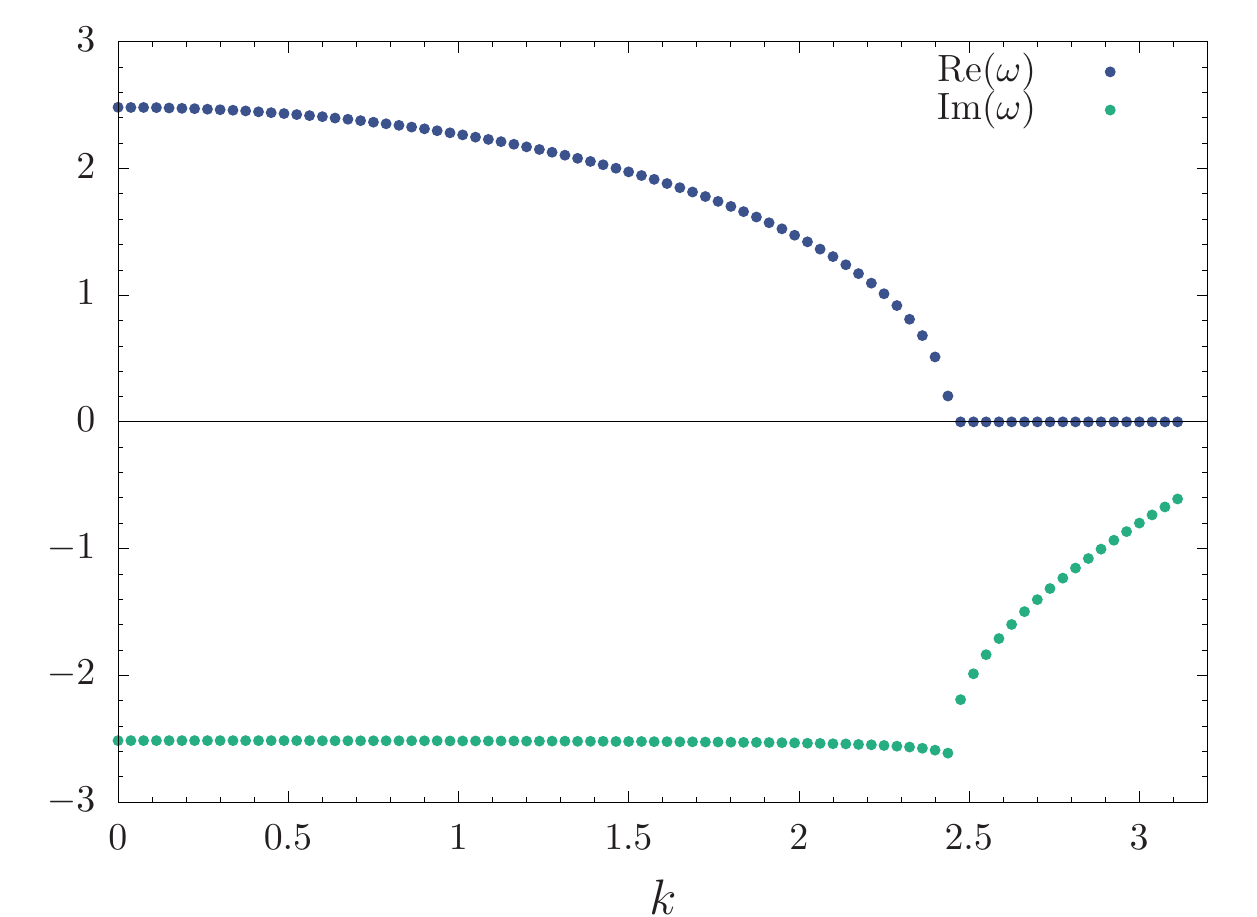}\\\medskip
 \includegraphics[width=0.5\textwidth]{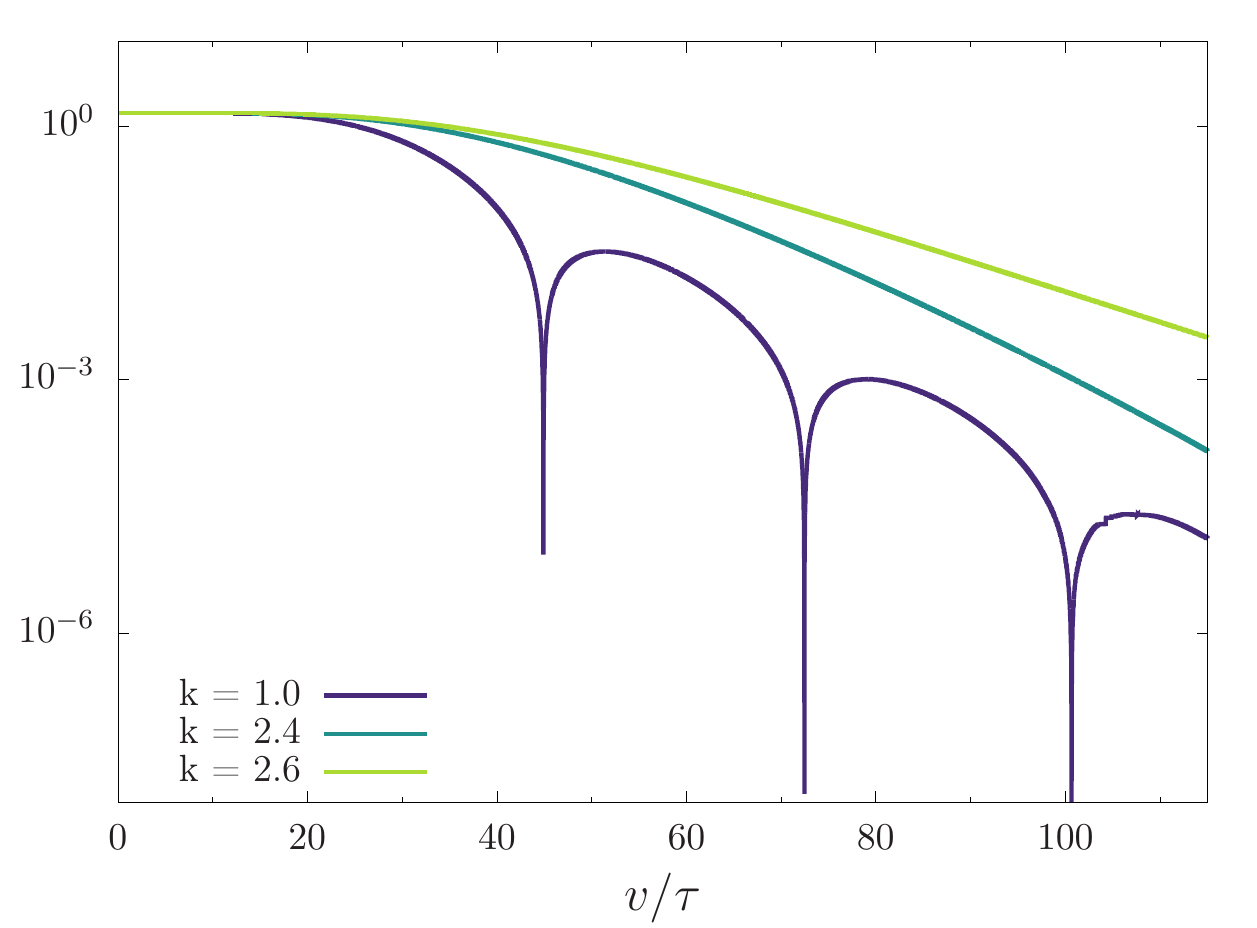}
 \end{center}
 \caption{The upper plot shows the lowest quasinormal mode in the final state for the case with vector magnetic field as a function of the momentum relaxation parameter $k$. A striking feature is that for large values of $k \gtrapprox 2.5$ the real
 part of the quasinormal mode is zero. The lower plot shows the same data as Fig. \ref{fig:BK} but on a logarithmic scale.
 We see nicely the periodic ringdown in the case of $k=1$ whereas the higher $k$ values show just exponential falloff as expected
 from the quasinormal mode analysis. Note also that the enveloping line for the curve with $k=1$ is practically parallel to the line with $k=2.4$ whereas the one with $k=2.6$ has a smaller slope.}
 \label{fig:BQNM}
\end{figure}

\begin{figure}
 \begin{center}
 \includegraphics[width=0.5\textwidth]{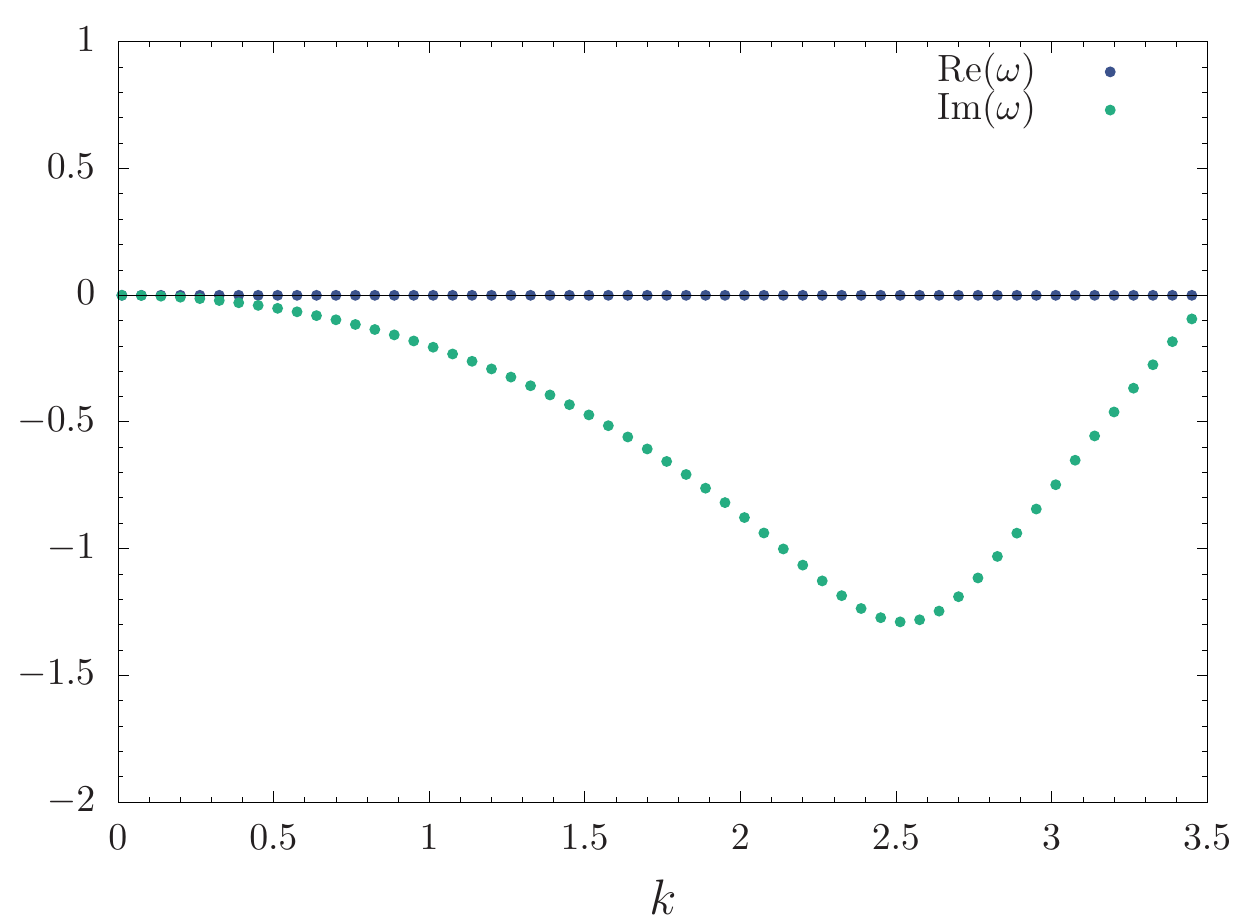}
 \\\medskip
 \includegraphics[width=0.5\textwidth]{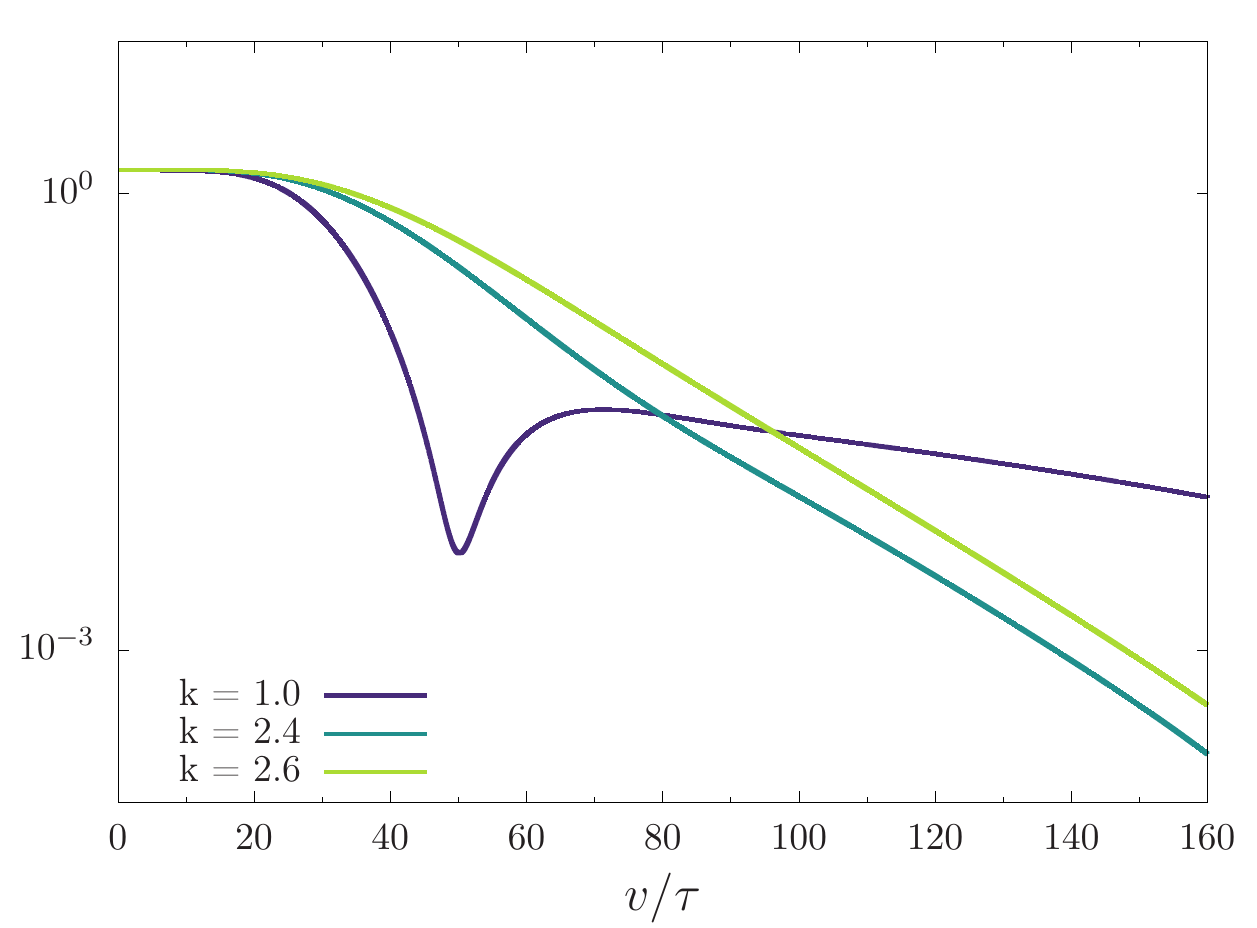}
 \end{center}
 \caption{The upper plot shows the lowest quasinormal mode in the final state for the case with axial magnetic field.
 This case is always dominated by a purely imaginary mode. Accordingly, the lower logarithmic plot does not show any sign of quasiperiodic ringdown. The slope becomes steeper for large $k$ values but stops doing so at around $k\approx 2.5$. This explains why the lines in the logarithmic plot for $k=2.4$ and $k=2.6$ are almost parallel. The datasets are the same as the ones in Figure \ref{fig:B5KJ}. We note that for early times and for $k=1$ the overshooting phenomenon is also seen nicely. In contrast the late time behavior is of course dominated by the lowest quasinormal mode.}
 \label{fig:B5QNM}
\end{figure}

We limit ourselves to the dominant mode with largest imaginary part for both cases. 
The real and imaginary part of the first quasinormal mode of the final equilibrium state are shown as functions of $k$ in Fig. \ref{fig:BQNM} for the case with vector magnetic field and in Fig. \ref{fig:B5QNM} for the case with axial magnetic field. It  can help us understand some of the features of the linear response results. In particular in what refers to the time of final equilibration of the currents, since the background will be equilibrated to this final equilibrium state before the current equilibrates.

For the case with vector magnetic field, the real part decreases while the imaginary part stays the same until the mode becomes purely imaginary slightly below $k=2.5$. At precisely that point and after increasing a little bit, the absolute value of the imaginary part decreases quite fast. This is in agreement with the equilibration times of Fig. \ref{fig:BK}. In that plot the final equilibration seems to happen roughly at the same point for $k$ below $2.4$, while for the curves above this value of the momentum relaxation parameter $k$ the equilibration time becomes longer and longer.
Since the quasinormal modes come in pairs with opposite-sign real parts, there there are two modes colliding on the imaginary
axes. The second mode then moves down the imaginary axes until in pairs up with another pure imaginary mode. 
In general a quite complicated pattern of modes moving on and off the imaginary axes develops for higher $k$ values.
Since our interest is in the dominating mode, we have not further investigated this.

For the case with axial magnetic field, the real part of the dominating mode is always zero, 
and therefore the mode is purely imaginary. 
One way of understanding this mode is that it is  connected to the diffusive mode at $k=0$. Since the diffusive mode
is purely imaginary, also this mode has a vanishing real part. 
The absolute value of the imaginary part first increases from zero, and after peaking roughly around $k = 2.5$, it decreases again to approach zero toward extremality, which is slightly above $k = 3.5$. This again shows good qualitative agreement with the energy current results of Fig. \ref{fig:B5KT}. We can see there that the curves approach the final value faster as $k$ grows but after $k = 2.4$ this trend changes and, in fact, the case with $k = 2.6$ equilibrates later.

\section{Conclusions}\label{sec:discussion}
We have studied the out-of-equilibrium chiral magnetic effect in a holographic setup based on the Vaidya type of solutions.
An essential and new ingredient is the inclusion of momentum relaxation. Without it, the response in the energy-momentum tensor would be trivial. 

A striking feature is that although the metric and gauge field backgrounds formally show an instantaneous equilibration, the anomaly-induced response shows a late equilibration. That proves that even in Vaidya metrics 
far-from-equilibrium linear response takes place. For sufficiently fast quenches the response is  delayed.
This means  the current does need some time to build up and it even stays at its original equilibrium value for a rather long time. 
Qualitatively, this is very similar to what has been observed in the previous study \cite{Landsteiner:2017lwm} based
on gravitational anomaly-induced chiral transport. Therefore, one might speculate that it is a universal feature of anomalous
transport. If so, it could hold some important lesson for the observation of the CME in heavy ion collisions at the RHIC and LHC.
While thermalization (or hydrodynamization) in general is believed to be very fast this might not necessarily mean that
the anomaly-induced charge separation sets in at the same timescale. There might be significant delay. Since the lifetime
of the magnetic field is smaller at the LHC it could mean that the chances for observing CME signals at the LHC are smaller than
those at the RHIC. One important direction for future research is therefore to develop better holographic out-of-equilibrium models of anomalous transport which take the finite lifetime of the magnetic field into account.

An interesting side result of our study is that the Vaidya metric allows for straightforward inclusion of momentum relaxation
with massless scalar fields. A curious feature is that the response in the nonconserved energy current builds up faster
for some intermediate value of the momentum relaxation parameter $k\approx 2.5$. This can be explained by looking to the
quasinormal mode spectrum. The value of the dominating mode first decreases, finds a minimum at around $k=2.5$ and then
starts to increase again. This indicates that for values $k>2.5$ the system relaxes more slowly to equilibrium and indeed 
we can see this also in the data for the time evolution of the energy current. 

There are many possible directions for future studies. For example the present work
could be generalized to out-of-equilibrium chiral vortical effect or nonlinear magnetic field dependence.
In view of the recent results of Ref.\cite{Cartwright:2019opv} on isotropization and thermalization in magnetic fields
it should also be rather interesting to study the CME in these types of far-from-equilibrium backgrounds.
Another question is if similar studies can be done with a dissipative conductivity, such as the electric conductivity.
Of course, one has to take into account that due to Joule heating an electric field will by itself introduce energy into the
system. It might be useful to investigate this in the decoupling limit in which this effect is suppressed.
Possibly the most important next step would be to include the finite lifetime of magnetic field and to choose regions of parameter space, charge, axial charge and temperature relevant
for modeling the quark gluon plasma.

\section{Acknowledgements}
The authors would like to thank Esperanza L\'opez, Christian Copetti, Matthias Kaminski and Casey Cartwright for very useful discussions. J.F.-P. would like to thank Javier Rodr\'iguez-Laguna and Thomas Biek\"otter for discussions on the numerical implementation. This work is supported by Grants No. SEV-2016-0597, FPA2015-65480-P and No. PGC2018-095976-B-C21 from MCIU/AEI/FEDER, UE. The work of J.F.-P. is supported by fellowship Grant. No. SEV-2012-0249-03. 

\bigskip

\appendix

\section{Numerical methods}\label{sec:appendix}

We used the fourth-order Runge-Kutta method both for the time and the radial numerical integration. 
The radial
derivatives of the fields were approximated using finite differences. 
In particular, we used fourth-order centered finite differences, except in the first two points near the boundary and the last two points beyond the horizon, where we used second-order finite differences, in the appropriate combination of centered, forward or backward. 

We used an initial spatial grid of 1000 (vector current) or 10,000 (axial and energy current) equally distributed points between $u = 0$ and $u = 1$.
This grid is supplemented with ten extra points inside the horizon. Along the time evolution the integration range 
is cut to ten points inside the apparent black hole horizon. The time step was taken to be at least 1 order of magnitude smaller than the radial step in order to guarantee stability of the algorithm. 

We saw that the computations were very sensitive to the boundary data and treating the first points right was the most important part of the numerical procedure. This is to be expected since the boundary of AdS is a regular singular point. To circumvent the associated complications we explicitly imposed in our equations the limit when $u$ goes to zero, using in some cases L'H\^{o}pital's rule with the asymptotic expansion, as we will see below.

We did not want the operators to be sourced, so we fixed the non-normalizable modes of all the different fields to zero. 
The first terms of the rest of the expansion, from the normalizable mode on, read
\begin{align}
 V =& V_2 u^2 + \dot{V_2} u^3 + {\cal O}(u^4) \, , \\
 A =& A_2 u^2 + \dot{A_2} u^3 + {\cal O}(u^4) \, , \\
 h =& h_4 u^4 - \frac{4 k Z_4}{5} u^5 + {\cal O}(u^6) \, , \\
 Z =& Z_4 u^4 + \frac{5 \dot{Z_4} - k h_4}{5} u^5 + {\cal O}(u^6) \, ,
\end{align}
where all the coefficients were functions of time only and $h_4$ had to satisfy
\begin{equation}
 \dot{h_4} = - k Z_4 \, .
\end{equation}
This constraint is clearly related to the conservation of $T^{0i}$ we already discussed in the main text, which is broken for $k\neq0$.

We can now substitute these asymptotic solutions in the definitions of the currents [Eqs.\eqref{currents:vector}, \eqref{currents:axial}, and \eqref{currents:tensor}] to obtain the responses parallel to the magnetic fields, which read
\begin{align}
 \langle J^z \rangle &= 2 V_2 \, , \\
 \langle J_5^z \rangle &= 2 A_2 \, , \\
 \langle T^{0z} \rangle &= 4 h_4 \, .
\end{align}
Therefore, obtaining our results boils down to extracting the normalizable modes. To do that we perform a least-squares fit according to the series expansion and read from the normalizable modes of the fields the one-point functions of the gauge and energy currents. We also use the vanishing of the coefficients of previous powers as a check for the accuracy of the method.

As we advanced above, we also used these series expansions in the equations of motion, substituting in $V$, $dV$, $A$, $dA$, $h$, $dh$, $Z$ and $dZ$, to obtain the form of the equations in the first point of the grid. After applying l'H\^{o}pital's rule to take the limit $u \to 0$, it can be seen that the radial equations reduce at that point to
\begin{align}
 \lim_{u \to 0}(\mathrm{eq.}\ref{Veqn}) &= 0 = dV' + V_2 \, , \\
 \lim_{u \to 0}(\mathrm{eq.}\ref{Aeqn}) &= 0 = dA' + A_2 \, , \\
 \lim_{u \to 0}(\mathrm{eq.}\ref{heqn}) &= 0 = dh' \, , \\
 \lim_{u \to 0}(\mathrm{eq.}\ref{Zeqn}) &= 0 = dZ' \, .
\end{align}
To proceed with the integration algorithm, we used in these equations the $V_2$ and $A_2$ that had been obtained in the fit from the previous time step. For the rest of the points in the grid, we did not need to use any information from the fit, only the numerical results of the fields and their derivatives.

Finally, to obtain the initial conditions, we assumed that the system started in equilibrium and obtained the equilibrium solutions for the appropriate values of $m$, $q_5$ and $k$. These equilibrium solutions could be found analytically solving the static version of the equations of motion, in which all the time derivatives were dropped. Their expressions are very lengthy and cumbersome. Since we do not consider them to be particularly informative, we have decided not to include them here. Our late time state will also be an equilibrium configuration and it can be checked that our latetime solutions approach the appropriate solutions of the static equations.

\bibliography{AnomTrans}{}

\end{document}